\documentclass[10pt,conference]{IEEEtran}
\IEEEoverridecommandlockouts

\usepackage{bbding}

\usepackage{graphicx}
\usepackage{subcaption}

\usepackage{anyfontsize}

\usepackage[dvipsnames]{xcolor}
\usepackage[many]{tcolorbox}

\usepackage{mathtools,amssymb,latexsym,amsfonts,stmaryrd}
\usepackage{pbox}
\usepackage{xfrac}
\usepackage{booktabs}
\usepackage{colortbl}
\usepackage{threeparttable}
\usepackage{amsthm}
\usepackage{scalerel}
\usepackage{multirow}
\usepackage{tikz}
\usepackage{mathrsfs}
\usepackage{mathpartir}
\usepackage{caption}
\usepackage{algorithm}
\usepackage{url}
\usepackage{syntax}
\usepackage{framed}
\usepackage[noend]{algpseudocode}
\usepackage{flushend}
\usepackage{listings}
\usepackage{moresize}
\usepackage{wrapfig}
 
\usepackage{seqsplit}
\usepackage{alltt}
\usepackage{xspace}
\usepackage{soul}
\usepackage{bm}
\usepackage{enumitem}
\usepackage{pifont}%

\newcommand{\reffig}[2]{\F~\hyperref[#1]{#2}}

\newcommand{\code}[1]{\textcolor{blue}{\textit{\bfseries{#1}}}}

\DeclareMathAlphabet{\mathcal}{OMS}{cmsy}{m}{n}

\DeclareMathOperator*{\argmin}{arg\,min} %

\usepackage{amsmath, listings, amsthm, amssymb, proof, xspace}

\usepackage{array}
\newcommand{\PreserveBackslash}[1]{\let\temp=\\#1\let\\=\temp}
\newcolumntype{C}[1]{>{\PreserveBackslash\centering}p{#1}}
\newcolumntype{R}[1]{>{\PreserveBackslash\raggedleft}p{#1}}
\newcolumntype{L}[1]{>{\PreserveBackslash\raggedright}p{#1}}

\usepackage{thmtools}

\declaretheoremstyle[spaceabove=\topsep,notefont=\normalfont\itshape]{mystyle}

\usepackage{footnote}
\makesavenoteenv{tabular}
\makesavenoteenv{table}

\usetikzlibrary{arrows,patterns, decorations.pathreplacing}
\newtheorem{definition}{Definition}

\newcommand{\F}{Fig.}
\newcommand{\E}{Eq.}

\definecolor{ForestGreen}{RGB}{34,139,34}

\newcommand{\parh}[1]{\noindent\textbf{#1}}
\newcommand{\parhs}[1]{\noindent\underline{\textit{#1}}}

\newcommand{\ignore}[1]{}

\lstdefinestyle{base}{
  moredelim=**[is][\color{red}]{@}{@},
  escapeinside={<@}{@>}
}

\newcommand{\tool}{\textsc{PhyFu}}
\newcommand{\eng}{\textsc{PSEs}}
\newcommand{\seng}{\textsc{PSE}}

\usepackage{amssymb}%
\usepackage{pifont}%

\newcommand\DejaVuttfamily{%
  \fontfamily{DejaVuSansMono-TLF}\selectfont }

\lstdefinestyle{base}{
  moredelim=**[is][\color{red}]{@}{@},
  escapeinside={<@}{@>}
}

\lstdefinelanguage
   [x64]{Assembler}     %
   [x86masm]{Assembler} %
   {morekeywords={CDQE,CQO,CMPSQ,CMPXCHG16B,JRCXZ,LODSQ,MOVSXD, %
                  POPFQ,PUSHFQ,SCASQ,STOSQ,IRETQ,RDTSCP,SWAPGS, %
                  rax,rdx,rcx,rbx,rsi,rdi,rsp,rbp, %
                  r8,r8d,r8w,r8b,r9,r9d,r9w,r9b}} %

\renewcommand{\baselinestretch}{.99}

\usepackage[compact]{titlesec}

\usepackage{listings}
\usepackage{fancyvrb}
\usepackage{color}
\definecolor{lightgray}{rgb}{.9,.9,.9}
\definecolor{darkgray}{rgb}{.4,.4,.4}
\definecolor{purple}{rgb}{0.65, 0.12, 0.82}
\definecolor{commentgreen}{RGB}{63,127,95}

\colorlet{myPurple}{blue!40!red}
\definecolor{myOrange}{RGB}{255,192,0}

\newcommand{\enc}[1]{$\phi^{*}_{\theta}$}
\newcommand{\dec}[1]{$\psi^{*}_{\theta}$}

\lstdefinelanguage{Solidity}{
  keywords={len,delete,int,void,payable, public, event, contract, typeof, new, true, false, catch, function, return, null, catch, switch, var, if, in, while, do, else, case, break,struct,const,socklen_t,sa_familty_t,char,sockaddr},
  keywordstyle=\color{violet}\bfseries,
  ndkeywords={class, export, boolean, throw, implements, import, this},
  ndkeywordstyle=\color{darkgray}\bfseries,
  identifierstyle=\color{black},
  sensitive=false,
  comment=[l]{//},
  escapeinside={(*@}{@*)},          %
  morecomment=[s]{/*}{*/},
  commentstyle=\color{commentgreen}\ttfamily,
  stringstyle=\color{red}\ttfamily,
  morestring=[b]',
  morestring=[b]"
}

\newcommand{\rnum}[1]{\uppercase\expandafter{\romannumeral #1\relax}}

\algnewcommand{\LeftComment}[1]{\Statex \(\triangleright\) #1}

\definecolor{pptbrown}{RGB}{132,60,12}
\definecolor{pptgreen}{RGB}{56,87,35}

\let\OLDthebibliography\thebibliography
\renewcommand\thebibliography[1]{
  \OLDthebibliography{#1}
  \setlength{\parskip}{0pt}
  \setlength{\itemsep}{0pt plus 0.1ex}
}

\definecolor{pptgreen}{RGB}{84,130,53}
\definecolor{pptred}{RGB}{176,35,24}

\definecolor{pptgreen1}{RGB}{78,173,91}
\definecolor{pptred1}{RGB}{192,0,0}

\definecolor{pptyellow1}{RGB}{203,195,167}
\definecolor{pptgreen2}{RGB}{184,192,176}

\definecolor{pptred3}{RGB}{192,0,0}
\definecolor{pptyellow3}{RGB}{255,192,0}
\definecolor{pptgreen3}{RGB}{4,216,178}

\definecolor{pptblue}{RGB}{0,176,240}
\definecolor{pptgrey}{RGB}{175,171,171}

\newlength{\dpcircle}
\newlength{\rcircle}
\newlength{\dcircle}

  \DeclareFontFamily{U}{dutchcal}{\skewchar \font =45}
  \DeclareFontShape{U}{dutchcal}{m}{n}{
    <-> dutchcal-r}{}
  \DeclareFontShape{U}{dutchcal}{b}{n}{
    <-> dutchcal-b}{}
  \DeclareMathAlphabet{\mdutchcal}{U}{dutchcal}{m}{n}
  \SetMathAlphabet{\mdutchcal}{bold}{U}{dutchcal}{b}{n}
  \DeclareMathAlphabet{\mdutchbcal} {U}{dutchcal}{b}{n}

  \DeclareFontFamily{U}{txcal}{\skewchar \font =45}
  \DeclareFontShape{U}{txcal}{m}{n}{
    <-> txr-cal}{}
  \DeclareFontShape{U}{txcal}{b}{n}{
    <-> txb-cal}{}
  \DeclareMathAlphabet{\mtxcal}{U}{txcal}{m}{n}
  \SetMathAlphabet{\mtxcal}{bold}{U}{txcal}{b}{n}
  \DeclareMathAlphabet{\mtxbcal} {U}{txcal}{b}{n}

\makeatletter
\newcommand{\algorithmicbreak}{\textbf{break}}
\newcommand{\BREAK}{\State \algorithmicbreak}
\makeatother

\usepackage[colorlinks]{hyperref}   %
\AtBeginDocument{%
  \hypersetup{pdfborder={0 0 1}, urlcolor=.}  %
}

\usepackage[capitalise]{cleveref}

\definecolor[named]{ACMBlue}{cmyk}{1,0.1,0,0.1}
\definecolor[named]{ACMYellow}{cmyk}{0,0.16,1,0}
\definecolor[named]{ACMOrange}{cmyk}{0,0.42,1,0.01}
\definecolor[named]{ACMRed}{cmyk}{0,0.90,0.86,0}
\definecolor[named]{ACMLightBlue}{cmyk}{0.49,0.01,0,0}
\definecolor[named]{ACMGreen}{cmyk}{0.20,0,1,0.19}
\definecolor[named]{ACMPurple}{cmyk}{0.55,1,0,0.15}
\definecolor[named]{ACMDarkBlue}{cmyk}{1,0.58,0,0.21}
\hypersetup{
  colorlinks,
  linkcolor=ACMPurple,
  citecolor=ACMPurple,
  urlcolor=ACMDarkBlue,
  filecolor=ACMDarkBlue
}

\usepackage{inconsolata} %
\usepackage{biolinum} %

\usepackage{microtype}

\usepackage[noadjust]{cite}

\newtheorem{example}{Example}

\newtheorem{theorem}{Theorem}

\def\BibTeX{{\rm B\kern-.05em{\sc i\kern-.025em b}\kern-.08em
    T\kern-.1667em\lower.7ex\hbox{E}\kern-.125emX}}

\begin{document}

\title{\tool: Fuzzing Modern Physics Simulation Engines}

\author{
  \IEEEauthorblockN{Dongwei Xiao, Zhibo Liu, and Shuai Wang\IEEEauthorrefmark{1}\thanks{\IEEEauthorrefmark{1} Corresponding author}}
  \IEEEauthorblockA{The Hong Kong University of Science and Technology,
  Hong Kong, China \\
  \tt \{dxiaoad, zliudc, shuaiw\}@cse.ust.hk}
}

\maketitle

\begin{abstract}

  A physical simulation engine (\seng) is a software system that simulates
physical environments and objects. Modern \eng\ feature both forward and
  backward simulations, where the forward phase predicts the behavior of a simulated system, and the backward phase provides gradients (guidance) for learning-based control tasks, such as a robot arm learning to fetch items.
  This way, modern \eng\ show promising support for learning-based control methods.
To date, \eng\ have been
largely used in various high-profitable, commercial applications, such as games,
movies, virtual reality (VR), and robotics. 
Despite the prosperous development and usage of \eng\ by academia and industrial
manufacturers such as Google and NVIDIA, \eng\ may produce incorrect
simulations, which may lead to negative results, from poor user experience in
entertainment to accidents in robotics-involved manufacturing and surgical
operations.

This paper introduces \tool, a fuzzing framework designed specifically for \eng\
to uncover errors in both forward and backward simulation phases. \tool\ mutates
initial states and asserts if the \seng\ under test behaves consistently with respect to basic
Physics Laws (PLs). We further use feedback-driven test input scheduling to
guide and accelerate the search for errors.
Our study of four \eng\ covers mainstream industrial vendors (Google and NVIDIA)
as well as academic products. We successfully uncover over 5K error-triggering
inputs that generate incorrect simulation results spanning across the whole
software stack of \eng.

\end{abstract}

\section{Introduction}
\label{sec:introduction}

Physics simulation engines (\eng) are computer software that simulate behavior of physical systems, such as rigid body dynamics
(e.g., a steel robot arm), soft body dynamics (e.g., elastic objects), and
fluid dynamics (e.g., water simulation). The past few decades have
witnessed a boom in using \eng\ in production environments, including computer
graphics, gaming, virtual reality (VR), and various robotic tasks like robot
control, robot parameter design, and trajectory optimization.

Physics simulation techniques have been studied for decades, with numerous
high-quality simulation engines developed and commercialized~\cite{Webots,
1389727, 6696520, 6094829, 10.1145/2776880.2792704, havok, ode,
physx,todorov2012mujoco}.
Notably, conventional \eng\ primarily aim for forward simulation, which
progressively computes the behavior of the simulated physical system starting
from an initial state. In contrast, modern \eng, often referred to as
``\textit{differentiable physical simulation engines},'' compose both forward and
backward simulation phases. While the forward phase still predicts how the
simulated system evolves, the defining characteristic of modern \eng\ is to
offer \textit{analytical gradients} via a backward phase. The ability of
computing analytical gradients makes it possible to perform end-to-end
optimization on agent control, and can speedup the optimization process by dozens
to hundreds of times~\cite{hu2019difftaichi}. Also, the simulated environments
become differentiable, making them technically compatible with training machine
learning-based control agents.
With the offered gradients, modern \eng\ have been demonstrated to
accelerate robotics control optimizations by one to four orders of magnitude
compared to conventional \eng~\cite{hu2019difftaichi}.
The market for \eng\ is also on a rapid rise, with both industrial and academic
efforts in developing and enhancing \eng~\cite{howell2022dojo, de2018end,
geilinger2020add,
werling2021fast,huang2021plasticinelab,freeman2021brax,heiden2021neuralsim}. The
total market share of developing and using \eng\ has reached over 10 billion
dollars and is expected to grow by over a 10\%
annually~\cite{meta-news,emergen-news,modor-news}. In addition, \eng\ can be
highly expensive, with certain surgical \eng\ licenses costing over over 10K
USD~\cite{comp-sur-sim}.

Nevertheless, production \eng\ often comprise dozens to hundreds of thousands of
lines of code~\cite{taichi-github,werling2021fast,warp,geilinger2020add},
covering a deep \textit{software stack}, including the simulation
algorithms~\cite{huang2021plasticinelab,heiden2021neuralsim}, hardware acceleration modules~\cite{taichi-github,warp,freeman2021brax}, even sometimes with domain
specific language (DSL) that has high expressiveness and efficiency on the
simulation primitives, as well as accompanying DSL compilers~\cite{taichi-github,warp}. Also, the simulated physical effects
considered by simulators are complicated, including collision detection,
friction, soft body dynamics, and fluid dynamics.
Plus, the derivation of gradients in the backward simulation is also
challenging, some requiring manual implementation of gradient computation~\cite{de2018end}, others requiring compilers to generate code for gradient computation directly from the forward simulation code~\cite{hu2019taichi,warp}.
All of these aspects make \eng\ highly complex systems demanding careful design
consideration of the \seng\ vendors. Notably, since \eng\ have been
applied in various scenarios, from the entertainment industry to safety-critical
sectors like surgery
robotics~\cite{davinci-sim,dv-trainer,robotix-mentor}, bugs in \eng, in
turn, can potentially lead to poor user experience or even catastrophic
accidents.

This work presents \tool, the first automated, systematic fuzz testing framework
for modern \eng. \tool\ tackles black-box scenarios, allowing testing of
(commercial) off-the-shelf \eng\ and holistically uncovering bugs in the full
software stack of \eng. Instead of capturing obvious ``crash''
behaviors (which are rare in \eng), \tool\ uncovers incorrect simulation outputs
(logic bugs) residing in both forward and backward simulation phases of \eng. To
this end, \tool\ generates and mutates the initial states of the system under
simulation, and it relies on a principled, clear testing oracle on the basis of
Physics Laws (PLs) to uncover incorrect simulations.

Moreover, although arbitrarily generating testing inputs can find a reasonable
number of errors during testing, such method struggles to generate inputs
towards regions that are difficult to reach, thus omitting some hidden bugs
or repeatedly exploring regions that are unlikely to trigger bugs. \tool,
instead, employs testing feedback to schedule the inputs and to drive the
fuzzing exploration toward space with a higher probability of violating our
oracle. Besides, the generated states must satisfy the real-world physical
constraints to ensure the validity of our testing inputs. It is seen that
randomly generating seed inputs can easily break the physical constraints, e.g.,
a robot arm penetrating the rigid ground. Instead of directly checking the
validity of testing inputs (a generally hard process), we generate testing seeds
by mutating an initially valid seed while ensuring the validity of our mutations
on the fly.

We implement \tool\ targeting four state-of-the-art \eng.
Brax~\cite{freeman2021brax}, Warp~\cite{warp}, and
Taichi~\cite{hu2019difftaichi} are developed by leading industrial \eng\
vendors: Google, NVIDIA, and Taichi Graphics, respectively. We also evaluate
Nimble~\cite{werling2021fast}, an advanced, open-source \eng\ developed by
Stanford.
Our experiments cover 8 combinations of \eng\ and physical scenarios, including
simulation on balls, robot arms, and soft bodies. \tool\ generates 10K test
inputs on each tested setting, and during approximately 20 days of testing, we
detected 5,932 inputs that resulted in erroneously simulated results. While the
discovered error-triggering inputs do not directly crash \eng, they silently
lead to incorrect simulations. We also found over 20 inputs that triggered
crashes, indicating severe security issues like buffer overflow.
Through manual analysis and feedback from the \eng\ developers, we found that
our discovered errors are due to a wide spectrum of reasons spanning the whole
stack of \seng\ software. By the time of writing, two bugs have been promptly
fixed.
In sum, we make the following contributions:

\begin{itemize}[leftmargin=*,topsep=0pt,itemsep=0pt]
\item We target a crucial yet under-explored need to test \eng. We, for the
  first time, propose an automated, systematic fuzzing framework for \eng\ in
  black-box settings. 

\item \tool\ makes several novel and practical design considerations and
optimizations to boost the fuzzing process and uncover more bugs. Also, \tool\
incorporates a set of strategies to ensure the validity of testing inputs.

\item Our large-scale evaluation of four modern \eng\ subsumes different
practical scenarios. Under all conditions, \tool\ detects a substantial number
of errors. Further root-cause analysis reveals a diverse set of hidden bugs
distributed across the full software stack of \eng.
\end{itemize}

We release and will maintain the codebase of \tool\ at~\cite{artifact} to boost
future research. 

\section{Preliminary}
\label{sec:background}

\begin{figure}[!htbp]
    \centering
    \includegraphics[width=0.9\linewidth]{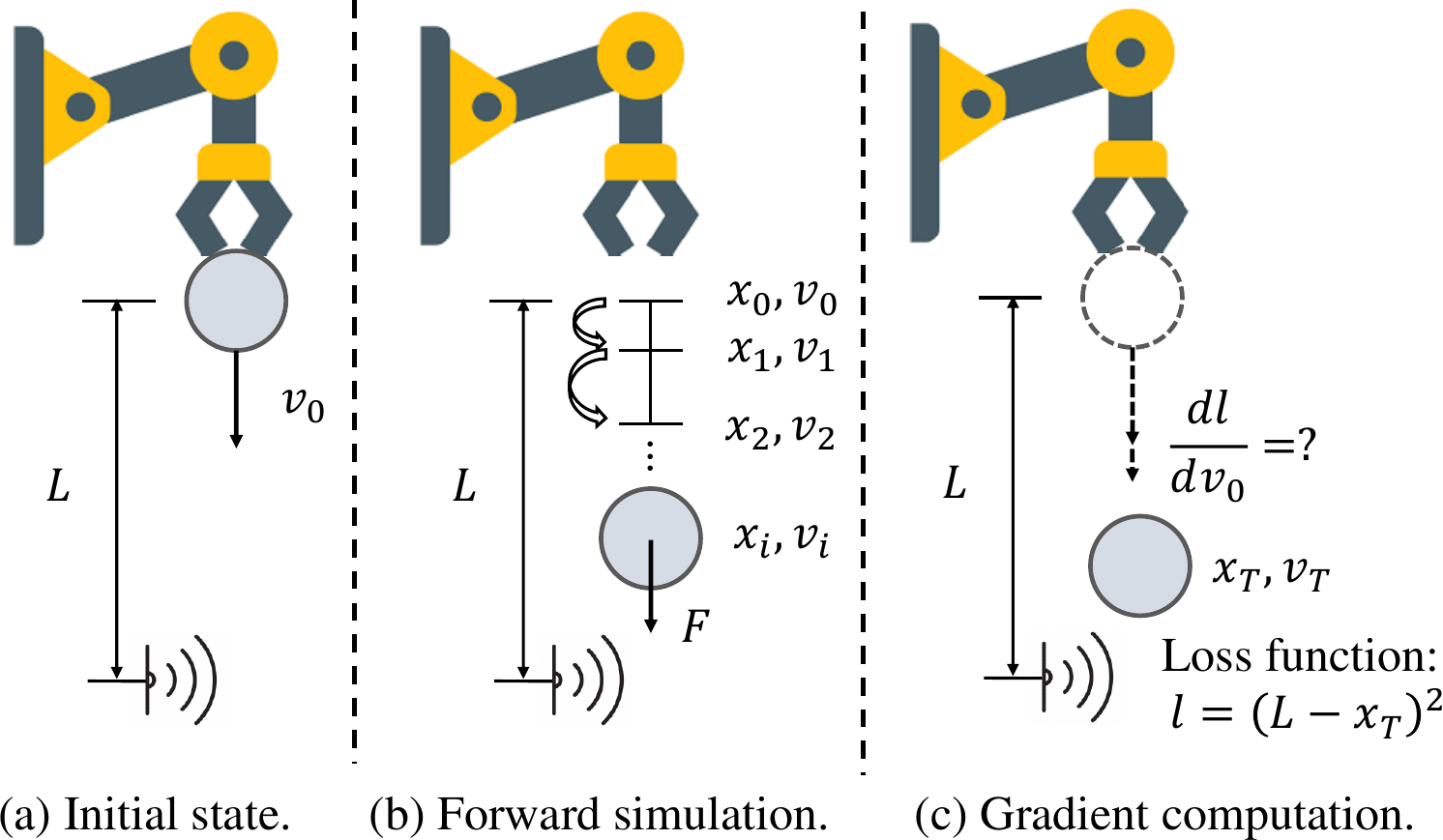}
    \caption{An example of the simulation process of modern \eng.}
    \label{fig:intro}
\end{figure}

This section introduces the preliminaries of \eng. For illustrative purposes, we
use a warehouse robotic arm as an example (\cref{fig:intro}).
\reffig{fig:intro}{1a} depicts the physical system in which a robot arm
releases a ball at initial speed $v_0$ downwards. The ball would perform
free-fall due to the gravitational force after the release. A sensor is located
$L$ meters below the ball's release point. The sensor would emit a probing
signal $ T $ seconds after the ball is released from the robot arm. The robot
aims to optimize the release speed of the ball so the sensor can sense the ball. In other words, the
learning goal of the robot is to control $ v_{0} $ to let the ball travel $ L $ meters in $ T $ seconds.

Modern \eng\ comprise two phases: forward simulation and backward simulation.
The forward simulation runs forward and predicts the system's behavior, while
the backward phase outputs analytical gradients that will be helpful for agent
learning and control. The details of the backward phase are more
intricate, and we introduce the forward simulation first.

\parh{Forward Simulation.}~The forward simulation models the position and
velocity of the objects in the physical system as a function of time $ t $. The
concatenation of position $ x $ and velocity $ v $ are refered to as
\textit{state}. \eng\ allow users to specify the \textit{external force},
denoted as $ \tau $, that is applied on the objects. The external force would
determine the change of states as a function of time.

Given the initial state $ s_{0} $ at $ t = 0 $ and the
external force $ \tau $ applied on the objects in the system, the
forward simulation phase predicts the state $ s(t) $ of the simulated system $
\mathcal{S} $ as a function of time $ t $ :
\begin{equation}
  \label{eq:abs-sim-func}
  s(t) = Sim(\mathcal{S}, s_{0}, \tau, t)
\end{equation}

Obtaining $ s(t) $ generally requires solving differential
equations deduced from physics laws, e.g., Newton's Second Law.

\begin{example}
  In the physical scenario of \cref{fig:intro}, the state of systems can be
  described by the dynamics equations in \cref{eq:newton-law} and
  \cref{eq:acceleration}:  
  \begin{equation}
     \label{eq:newton-law}
      F = mg = m \frac{dv}{dt}
  \end{equation}
  \begin{equation}
    \label{eq:acceleration}
    \frac{dx}{dt} = v
  \end{equation}
, where \cref{eq:newton-law} denotes that the ball would accelerate due to
gravity, according to Newton's Second Law, and \cref{eq:acceleration} comes from
the fact that the time-derivative of position is velocity.
\end{example}

To obtain the function $ s(t) $, \eng\ would discretize the time $ t $ into dozens
to hundreds of small intervals, each with length $ \Delta t $. Later, as
illustrated in \reffig{fig:intro}{1b}, \eng\ deduce the system state $
s_{i+1} $ at the time interval $ [i\Delta t, (i + 1)\Delta t) $ based on the
previous state $ s_{i} $. The derivation of $ s_{i + 1} $ from $ s_{i} $ is
determined by a \textit{time-stepping function} $ TS $:
\begin{equation}
    \label{eq:forward-seq}
    s_{i + 1} = TS(s_i, \tau_i)
\end{equation}
, where $ \tau_{i} $ is the external force applied on the objects at the given
time interval. $ TS $ is determined by the differential equations to be solved.

\parh{Backward Simulation.}~Learning and control tasks, e.g., a robot controlling
the initial velocity to shoot a basketball into the basket, typically aim at deciding an initial state $ s_{0} $ that would lead to the optimal objective
function $ h(s_{t}) $ defined on the final state $ s_{t} $. 
In general, the key requirement of the agent learning process is to enable access to the
analytical gradients, $ \partial h(s_{T}) / s_{0} $, of function $ h $ w.r.t.
the initial state $ s_{0} $, as this would enable an end-to-end optimization on the
objective function directly. It has been shown that analytical gradients can
speed up the optimization process by tens to hundreds of
times~\cite{hu2019taichi,freeman2021brax}.

As one key feature,
modern \eng\ offer the ability to compute analytical gradients in the backward simulation phase. This way, \eng\ provide feedback to guide learning-based agents, like robotics control models, to gradually improve their performance using gradients and reach a user-specified objective. In contrast to the forward
simulation, gradient computation runs backward, starting from a loss function
defined on the final state and propagating the gradients to the initial state.

Consider \reffig{fig:intro}{1c}, suppose the forward simulation
process determines that the ball would be located at $x_T$ after $T$
seconds, where $x_T \neq L$. Given an objective such that the ball is required
to be at location $L$ at time $ T $ when a probing signal is emitted, the robot arm learns to gradually adjust the initial speed, $v_0$, of
the ball. The optimization goal can be formulated as
\cref{eq:example-opt-goal}:
\begin{equation}
    \label{eq:example-opt-goal}
    \argmin_{v_0} \left| x_T - L \right|^2
\end{equation}

Modern \eng\ facilitate solving \E~\ref{eq:example-opt-goal} in a principled
manner by providing the gradients of the simulation process and making the
entire process \textit{differentiable}. Thus, the simulation process can be
smoothly integrated into training a learning-based agent.
\reffig{fig:intro}{1c} illustrates how \eng\ can be employed. Given the loss
function defined as $l = (L - x_T)^2$, the backpropagation first computes $dl /
dx_T$, which can be easily computed by automatic differentiation frameworks such
as PyTorch~\cite{NEURIPS2019-9015}. Then, \eng\ can derive the analytical
gradients $dx_T / dv_0$ (see below for details). The $ dl / dx_T $ and $ dx_T /
dv_0 $ can then be combined via chain rule, resulting in the end-to-end gradient
$dl / dv_0$, i.e., the gradient of the loss value $l$ w.r.t. the parameter
$v_0$. Having obtained the end-to-end gradients, parameters in a learning-based
agent like robot arm can be consequently optimized using methods like gradient
descent.

Various methods have been proposed for gradient derivation. Some engines
formulates the simulation process as a linear complementarity problem
(LCP)~\cite{werling2021fast,de2018end} and derive gradients accordingly, and
some treat the simulation as a non-linear complementarity problem
(NCP)~\cite{heiden2021neuralsim,howell2022dojo} that conforms to the physical
laws better but is more costly to solve. Some engines also use compliant
models~\cite{freeman2021brax,jatavallabhula2021gradsim}, convex optimization
models~\cite{zhong2021extending}, and position-based dynamics~\cite{warp}.

\section{Overview of \tool}
\label{sec:overview}

We aim to uncover logic bugs in \eng. As in \cref{sec:background}, \eng\ feature
both forward and backward phases to train learning-based complex
agents~\cite{hu2019difftaichi}. The forward phase computes the final state $
s_{T} $ of a physical system at time $ T $ for a given initial state $s_0$. The
backward, gradient-computation phase starts from a loss function defined on the
final state, and propagates the gradients backwards to the initial state.

To systematically subsume both forward and backward phases, we propose a testing
oracle based on principled Physics Laws (PLs). Before introducing the PLs
used, we first list some assumptions below:

\parh{Assumptions and Application Scope.}~We assume a set of reasonable
pre-conditions so that the PLs we use can hold. The assumptions and their
corresponding proofs are already well-researched in a category of
mathematics and physics problem called the ``Inverse
Problem''~\cite{collar2002uoso,hasanov2017ItIP}, which deals with whether the initial state $ s_{0} $ for an observed system's final state $ s_{T} $ is unique and proposes strategies to find such $ s_{0} $. Due to the complexity of the
strict definitions and proofs, we can only briefly list the high-level ideas of
the full assumption set in a less sound and complete form, so as to facilitate
the understanding of audience from general background:

\begin{enumerate}
  \item The physical process is deterministic and non-chaotic.
  \item The final state after the forward physical process has to depend
  continuously on the initial state.
  \item No friction is allowed in the physical system.
\end{enumerate}

We note that the above assumptions can reasonably hold in real-world physical
systems and \eng. The first two assumptions are easily satisfied in common use
cases of \eng, such as robotic and soft-body simulations. The third
assumption is a common setting in real-world usage of \eng, since typical
physical scenarios under simulation, such as rigid
simulation~\cite{hu2019taichi}, molecular
simulation~\cite{michaud2011mdanalysis,frenkel2001understanding} and fluid
simulation~\cite{muller2003particle,sadus2002molecular}, generally configure the
friction force to be zero, as the support for friction from \eng\ is not mature enough and still an open-research
problem~\cite{howell2022dojo,hu2019taichi,huang2021plasticinelab}.

We present the following important fact, denoting the uniqueness of $s_0$ (on
the condition of assumptions listed above): %

\begin{tcolorbox}[size=small]
  Under fixed external force, the $ s_0 $ that can lead to the observed $ s_T $ is \textit{unique}~\cite{MeshcheryakovG.A.1970Uots,KapanadzeD.V.2008Otuo,PrilepkoA.I.1971Otsa,RammA.G1980A1SS,collar2002uoso,hasanov2017ItIP}.
\end{tcolorbox}

\parh{Forward Testing Oracle.}~Based on the property of uniqueness for $ s_{0} $
in the theory above, we formulate our testing oracle to test the forward
simulation process as:

\begin{definition}[Forward Testing Oracle]
    \label{def:forward-oracle}
    Let $ f(s_0) = s_T$, where $ f $ is the forward simulation mapping initial
    state $ s_{0} $ to final state $ s_{T} $.\footnote{Since we are
    considering the relation between $ s_{0} $ and $ s_{T} $, the $ \mathcal{S}$,
    $\tau$, and $T $ in \cref{eq:abs-sim-func} are assumed to be all fixed and are
    left out in $ f $ for simplicity. The same applies to the rest of the
    paper.} For $\forall s_0'$ s.t. $f(s_0') = s_T$, we assert if $s_0' = s_0
    $.
\end{definition}

Oracle in \cref{def:forward-oracle} asserts that the final state should
\textit{not} be identical whenever two initial states are distinct. If this
property is not adhered to, it indicates the presence of bugs in the forward
simulation phase of the tested \eng.

\smallskip
\parh{Backward Testing Oracle.}~To test the analytical gradients yielded from
the backward simulation, we use a theoretic property of gradients:

\begin{theorem}[Theory of Gradient-Based Optimization]
  \label{theo:grad}

  Consider a differentiable multi-variate function $ h(x) $ and its global minimal point $ x^{*} $. Starting from a point $ x = x^{*} + \Delta x^{0} $ that lies sufficiently close to $ x^{*} $, perform gradient descend about $ \Delta x $ until convergence:
  \begin{align}
    \Delta x^{k+1} &= \Delta x^{k} - \alpha^{k} g(\Delta x^{k}) \label{eq:grad-descend}
  \end{align}
  , where $ k $ is the iteration number, and $ \alpha^{k} $ is the update step
  length. Then $ h(x) $ is bound to converge globally to $ h(x^{*})
  $~\cite{cottle2017linear}.
\end{theorem}

\begin{tcolorbox}[size=small]
\cref{theo:grad} implies that starting from a nearby point around the optimal
  point $ x^{*} $ of the objective function $ h(x) $, the gradient $ g(x) $
  should be able to guide the optimization of $ h(x) $ towards $ h(x^{*}) $
  with sufficient iterations.
\end{tcolorbox}

\begin{figure}[htbp]
  \centering
  \hfill
  \begin{subfigure}{0.48\linewidth}
      \centering
      \includegraphics[width=\linewidth]{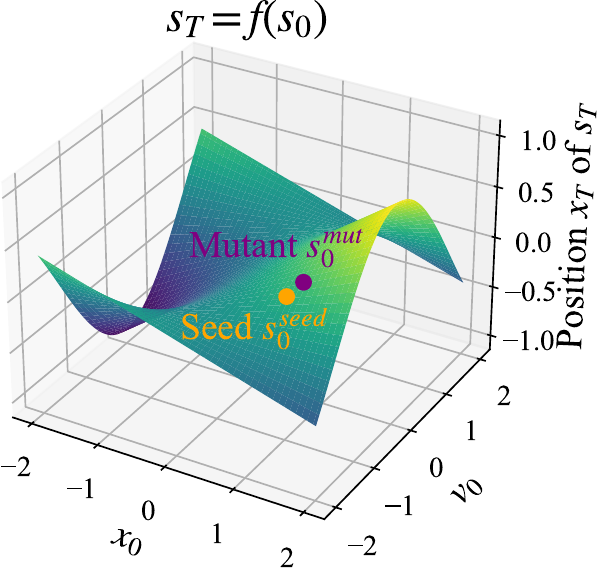}
      \caption{Seed and mutant.}
      \label{subfig:seed-mutated-init}
  \end{subfigure}
  \hfill
  \begin{subfigure}{0.48\linewidth}
      \centering
      \includegraphics[width=\linewidth]{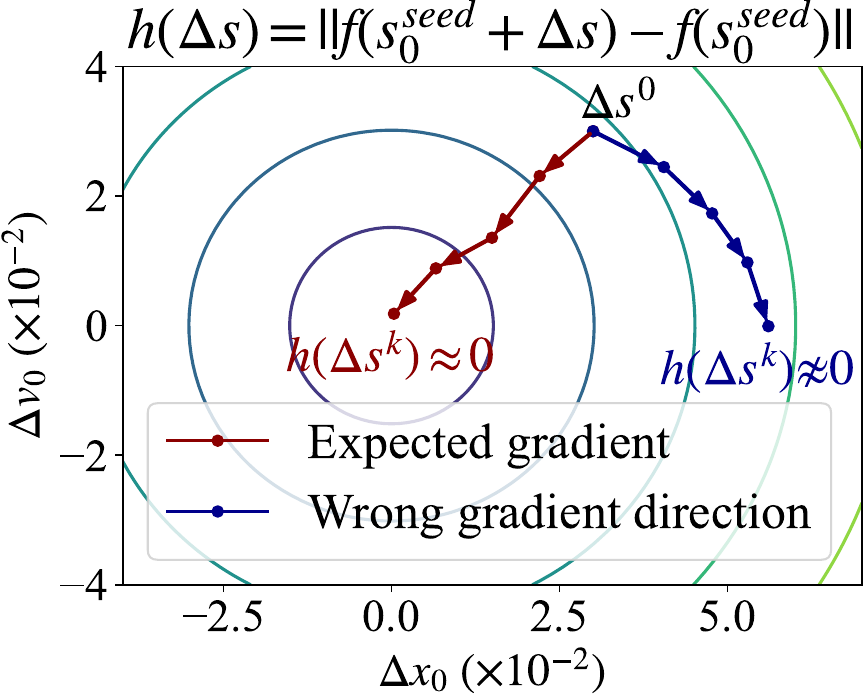}
      \caption{Optimize $ \Delta s $ with gradients.}
      \label{subfig:grad-desc}
  \end{subfigure}
  \hfill
  \caption{Illustration of the backward oracle. In (a), $ s_{0}^{mut} $ is
  obtained by adding a small $ \Delta s $ with $ s_{0}^{seed} $; (b) shows two traces of gradient descent iterations starting from the initial $ \Delta s
  $, where red and blue traces are guided by the correct and wrong gradients,
  respectively. The intuition of our backward oracle is reflected from the blue
  trace.}
  \label{fig:backward-example}
\end{figure}

We illustrate the ideas of our backward simulation testing approach in
\cref{fig:backward-example}. Denote the forward simulation function as $ f $,
which maps an initial system state $ s_{0} $ to the final state $ s_{T} $. In
\cref{subfig:seed-mutated-init}, the position $ x_{0} $ and velocity $ v_{0} $
component of initial state $ s_{0} $ are shown on the x-y plane, and the z-axis
shows the position component $ x_{T} $ of the corresponding final state $ s_{T}
$ ($ s_{T} = f(s_{0}) $). Given an initial state $ s_{0}^{seed} $ and its final
state $ s_{T}^{seed} $, we randomly add a \textit{small perturbation}, $ \Delta
s $, to $ s_{0}^{seed} $, and obtain a mutated initial state $ s_{0}^{mut} $. We
then aim to find the minimum of the objective function:
\begin{equation}
  \label{eq:backward-objective}
  h(\Delta s) = \lVert f(s_{0}^{seed} + \Delta s) - f(s_{0}^{seed}) \rVert
\end{equation}
, by executing the gradient-descend algorithm on $ \Delta s $ under the guidance
of gradients $ g(\Delta s) $ computed from the backward simulation.
\cref{subfig:grad-desc} shows two example optimization traces, both of which
start from the initial $ \Delta s^{0} $ and end in their respective $ \Delta
s^{k} $ after $ k $ gradient-descend iterations. The red trace shows the
optimization trace guided under the correct gradients, while the blue one is
guided by buggy gradients. After $ k $ iterations, the red one successfully
finds the minimal point, i.e., when $ h(\Delta s) \approx 0 $, while the blue one
causes $ h(\Delta s) $ to deviate from its minimum further and further
away due to the wrong direction of computed gradients. Based on
\cref{theo:grad}, we assert the following oracle to test backward simulation:

\begin{definition}[Backward Testing Oracle]
  \label{def:backward-oracle}
  Starting the gradient descend on a small initial $ \Delta s $, the
  objective function $ h(\Delta s) $ should converge to 0 after sufficient number of iterations.
\end{definition}

The proof of \cref{def:backward-oracle} can be readily derived from \cref{theo:grad}, and we leave the full proof on our website~\cite{artifact} due to space limit.

\begin{figure}[!htbp]
    \centering
    \includegraphics[width=1.0\linewidth]{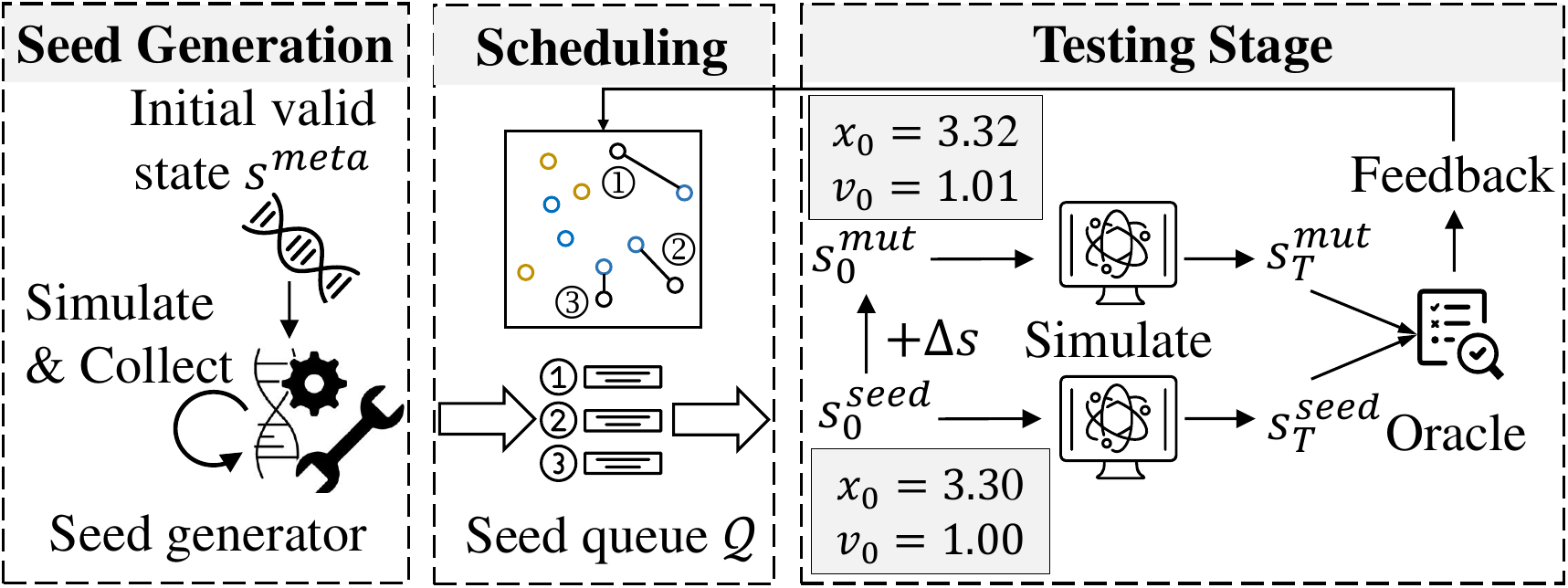}
    \caption{Testing pipeline.}
    \label{fig:pipeline}
\end{figure}

\section{Design of \tool}
\label{sec:design}

\parh{Design Overview.}~\tool\ delivers an automated, systematic fuzzing
framework to test modern \eng\ with respect to our forward and backward testing
oracles. 
\cref{fig:pipeline} illustrates the pipeline of \tool, which consists of three
parts: \ding{192} seed generation, \ding{193} seed scheduling, and \ding{194}
testing. 

\parhs{\ding{192} Seed Generation.}~\tool\ generates a pool of states $
\mathcal{C} $ as the testing seeds for the fuzzing. A state $ s^{seed} $ in the
seed pool $ \mathcal{C} $ comprises the position and the velocity of objects in
the system simulated by the \seng\ under test. The generated seed states will be
scheduled and fed into the testing stage for uncovering erroneous simulations.
We propose a set of design choices to ensure the validity of seed states during
the seed generation phase (see \cref{subsec:seed-validity}).

\parhs{\ding{193} Seed Scheduling.}~Testing by randomly generating seeds can
suffer from inefficiency and ineffectiveness issues. Since the simulation
process of \eng\ is highly time-consuming (often a testing campaign takes
several to dozens of hours; see \cref{tab:results}), to uncover more bugs with
a reasonable resource, we employ a scheduling algorithm to prioritize seeds that are more
likely to trigger bugs. The seeds generated from the seed generator are enqueued
and sorted in a seed queue $ \mathcal{Q} $, and are dequeued to be used in the
testing stage based on their priority.

\parhs{\ding{194} Testing.}~The testing stage dequeues a seed $ s^{seed} $
from the seed queue, and uses it as the initial state, $ s_{0}^{seed} $, for the
simulation. The initial state is added with a sufficiently small mutation, $
\Delta s $, to obtain a mutated initial state $ s_{0}^{mut} $. We will then
check the seed and mutant against our forward and backward oracles, and provide
testing outcomes as feedback to the seed scheduling stage to
prioritize unused seeds.

\subsection{Fuzz Testing}

\label{subsec:fuzzing}
\setlength{\textfloatsep}{10pt}
\begin{algorithm}[!t]
\caption{Fuzz Testing.}
    \footnotesize
\label{alg:fuzzing}
  \begin{algorithmic}[1]
\Function{\code{Fuzzing}}{Corpus of Seed Inputs $\mathcal{C}$, $T, \mathcal{E}$}
    \State $\mathcal{Q} \leftarrow \mathcal{C}, \mathcal{O} \leftarrow \varnothing, \mathcal{F} \leftarrow \varnothing $
    \For{1 ... \textit{MAX\_ITER}}
      \State $s_{0}^{seed} \leftarrow $ \textsc{Dequeue}($\mathcal{Q}$)
      \State $ \tau \leftarrow randForce() $ \Comment{Randomly apply external forces.}
      \State $s_{T}^{seed} \leftarrow $ \textsc{Forward}($\mathcal{E}, s_{0}^{seed}, \tau, T$)
      \State $ \Delta s \leftarrow \code{Mutate}( s_{0}^{seed} $)  \Comment{Generate small perturbation.}
      \For{1 ... $\textit{MAX\_GRAD\_ITER}$}
        \State $ s_{0}^{mut} \leftarrow s_{0}^{seed} + \Delta s $ \Comment{Add mutation.}
        \State $s_{T}^{mut} \leftarrow $ \textsc{Forward}($\mathcal{E}, s_{0}^{mut}, \tau, T$)
        \State $\mathcal{L} \leftarrow \lVert s_{T}^{seed} - s_{T}^{mut} \rVert^2 $ \Comment{Compute loss.}
        \If {$ \mathcal{L} < \epsilon_{B} $} 
          \BREAK
        \EndIf
        \State $ g \leftarrow \textsc{Backward}(\mathcal{E}, \mathcal{L}) $ \Comment{Backward simulation.}
        \State $\Delta s \leftarrow $ \code{Update}($\Delta s, g$) \Comment{Gradient descend.}
      \EndFor
      \If{ \code{ViolateOracle}($ s_{0}^{seed}, s_{0}^{mut}, s_{T}^{seed}, s_{T}^{mut}, \mathcal{L} $) }
        \State \textsc{Add}($\mathcal{O}, s_{0}^{seed}$) \Comment{Include in the buggy set.}
      \EndIf
      \State \textsc{Add}($ \mathcal{F}, s_{0}^{seed} $) \Comment{Mark as already-fuzzed.}
      \State \code{Schedule}($\mathcal{Q}, \mathcal{O}, \mathcal{F} $) \Comment{Seed scheduling.}
    \EndFor
    \State \Return Buggy states $\mathcal{O}$
    \EndFunction
  \end{algorithmic}
\end{algorithm}

\cref{alg:fuzzing} formalizes the fuzzing procedure, which tests a target \seng\
$\mathcal{E}$ and yields a set of error-triggering inputs $ \mathcal{O} $ that
violates our testing oracles, either \cref{def:forward-oracle} or
\cref{def:backward-oracle}. The core components of \cref{alg:fuzzing} can be
summarized below:

\begin{itemize}[leftmargin=*]
  \item \code{Mutate} takes the initial seed state $ s_{0}^{seed} $ as input and
  generates a small perturbation $ \Delta s $, which will later be added to $
  s_{0}^{seed} $ to obtain a mutated initial state $ s_{0}^{mut} $. The mutant $
  s_{0}^{mut} $ will be used to initiate the fuzzing campaign.

\item \code{Update} optimizes the amount of
perturbation, $ \Delta s $, based on its gradients $ g $ computed by the
backward simulation. The function applies a step of the gradient-descend algorithm,
as the one mentioned previously in \cref{eq:grad-descend}.
  
\item \code{ViolateOracle} checks whether our oracle,
  \cref{def:forward-oracle} or \cref{def:backward-oracle}, is violated. It takes
  as input the initial and final states of both the seed and the mutant, as well
  as the final loss value after the gradient descend, and checks against the
  forward and backward oracles.

  \item \code{Schedule} arranges the order of the seeds. The fuzzing seeds are
  stored and sorted in a queue $ \mathcal{Q} $, and are dequeued based on their
  priority. To maximize the efficiency of fuzzing and find more bugs in a
  limited time budget, we leverage a feedback-driven scheduling algorithm to
  prioritize the seeds (see \cref{subsec:scheduling}).
\end{itemize}

Our fuzzing algorithm comprises a series of carefully designed steps to trigger
simulation failures of the target \seng. Overall, \cref{alg:fuzzing} contains
the following steps:

\parh{Initialization.}~As the entry point of our procedure, \code{Fuzzing} takes
as input a corpus of seed inputs $\mathcal{C}$, a simulation time span $T$, and
a target PSE $\mathcal{E}$. 
$ \mathcal{C} $ is used to initialize a queue $  \mathcal{Q} $ (line 2), which
determines the order of seeds during fuzzing.

\parh{Fuzzing Loop.}~In each iteration, we pop out the seed with the highest
priority (see below for \code{Schedule}), and use it as the seed initial state $
s_{0}^{seed} $ (line 4), whose corresponding final state $ s_{T}^{seed} $ is
obtained from the forward simulation of $\mathcal{E}$ (line 6). Based on $ s_{0}^{seed} $, we generate a small perturbation amount $ \Delta s $ (line 7), which will later be added over the seed initial state $ s_{0}^{seed} $ to obtain a mutated initial state $ s_{0}^{mut} $ (line 9).

\parh{Gradient Descend Iterations.}~The gradient desend iterations (lines 8--15) follow the
steps introduced in \cref{sec:overview} to check the backward simulation.
Starting from the initial perturbation amount $ \Delta s $ generated in line 7, the variable $ \Delta s $ is iteratively updated under the guidance of its gradients, $ g $, from the
backward simulation. The iteration loop is terminated if the loss value $
\mathcal{L} $, which is defined as the distance between the final states $
s_{T}^{seed} $ and $ s_{T}^{mut} $, is small enough (line 12), or the number of
iterations reaches a pre-defined threshold $ MAX\_GRAD\_ITER $ (line 8).

\parh{Oracle Violation Checking.}~After the gradient descend iterations,
\code{ViolateOracle} detects any violation against the forward or backward
oracle.

\parhs{Backward Oracle Checking.}~The backward oracle checking examines the loss
value $ \mathcal{L} $, which is iteratively updated under the guidance of
gradients $ g $. If $ \mathcal{L} > \epsilon_{B} $, then the backward oracle,
\cref{def:backward-oracle}, is deemed to be violated, and gradients
from the backward simulation phase are considered buggy.

\parhs{Forward Oracle Checking.}~Following \cref{def:forward-oracle}, when the
final states of seed and mutant are close to each other, we check if their
corresponding initial states are close to each other. 
Specifically,
\code{ViolateOracle} would report oracle violation if $ \lVert s_{T}^{seed} -
s_{T}^{mut} \rVert < \epsilon_{F} $ (near-identical final state) while $ \lVert
s_{0}^{seed} - s_{0}^{mut} \rVert > \epsilon_{I} $ (distinct initial state).

\parh{Adaptive Fuzzing Scheduling.}~After each fuzzing iteration, $ s_{0}^{seed}
$ would be included in the output set $ \mathcal{O} $ if an oracle violation is
detected (line 17). The buggy set $ \mathcal{O} $, together with the set $
\mathcal{F} $ of all the already-used seeds, will be used as feedback to
prioritize the seeds in the seed queue $ \mathcal{Q} $ (line 19). Details of
prioritization are introduced in \cref{subsec:scheduling}.

\subsection{Feedback-Driven Seed Prioritization}
\label{subsec:scheduling}

\parh{Inefficiency of Random Seed Selection.}~The testing campaign is seen to be
slow in evaluations (see \cref{tab:results}). The forward simulation process
requires discretizing the simulation time span $ T $ into dozens to hundreds of
small steps (see \cref{sec:background}), and iteratively applying a non-trivial
time-stepping function \cref{eq:forward-seq} for each small step; the backward
simulation also consumes comparable time as the forward simulation. As such, the
testing process can take hours or dozens of hours to finish, spending a
considerable amount of time on the simulation process itself. Randomly selecting
a seed from the seed pool can be inefficient, as it can waste time executing the
costly simulation on seeds that are not likely to trigger errors.

\parh{Prioritization Strategy.}~To uncover more errors in a limited time budget,
we employ the Adaptive Random Test (ART) algorithm~\cite{chen2007test} to
prioritize seeds with higher probabilities to trigger errors. The intuition
behind ART is that neighbors of a non-failure-inducing seed are also less likely
to cause errors, while neighbors of a failure seed are also more likely to cause
failures. \cref{alg:feedback} shows the workflow of ART. Given the set of
error-triggering seeds $ \mathcal{O} $ and the set of already-used seeds $
\mathcal{F} $, ART would first derive non-failure-causing set $ \mathcal{U} $ by
taking the difference of $ \mathcal{F} $ and $ \mathcal{O} $ (line 2). For each
seed to schedule in the seed queue (line 4), it will compute the minimal
distance to all the non-failure-causing seeds (lines 5--7), and take the
computed minimal distance as the corresponding energy value (line 8). The seeds
in $ \mathcal{Q} $ will be sorted in descending order according to their energy
value (line 9).
ART allows us to focus our testing efforts on the most promising seeds,
increasing the chances of finding bugs and wasting less time exploring
uninteresting states; see \cref{subsec:ablation} for empirical results.

\begin{algorithm}[!t]
\caption{Feedback-Driven Test Case Prioritization}
    \footnotesize
\label{alg:feedback}
  \begin{algorithmic}[1]
\Function{\code{Schedule}}{Seed queue $\mathcal{Q}$, Buggy set $ \mathcal{O} $, Full set $ \mathcal{F} $}
    \State $ \mathcal{U} \leftarrow \mathcal{F} - \mathcal{O} $ \Comment{$ \mathcal{U} $ as seeds that don't trigger errors.}
    \State $ EnergyList \leftarrow zeros(\mathcal{Q}.size) $
    \For {$ i $ in range($ \mathcal{Q}.size $)}
      \State $ minDist \leftarrow +\infty $
      \For {$ u $ in $ \mathcal{U} $} \Comment{Find minimal distance to seed states in $ \mathcal{U} $.}
        \State $ minDist \leftarrow min(minDist, \lVert u - \mathcal{Q}[i] \rVert) $
      \EndFor
      \State $ EnergyList[i] \leftarrow minDist $
    \EndFor
    \State \textsc{Sort}($ \mathcal{Q} $, $ EnergyList $) \Comment{Sort in descending order.}
    \EndFunction
  \end{algorithmic}
\end{algorithm}

\subsection{Seed Validity Ensurance}
\label{subsec:seed-validity}

It is critical to ensure that our testing seeds, denoting initial states
$s_{0}^{seed}$ of the tested PSE, are valid in the physical world. Otherwise,
``bugs'' found by the testing process may be less interesting or even false positives.
Typically, the requirements for the seed validity can be encoded as state
constraints, or referred to as \textit{constraints}. Constraints vary from
scenario to scenario. We will use the example of two balls colliding with each
other to illustrate the concept of constraints:

\begin{example}
  \label{exam:two-balls}
  In a simulation involving only rigid objects, the states of two balls must
  satisfy a set of constraints, namely, ``intersection constraint''. A rigid
  object cannot penetrate another rigid one. In the situation of two balls, the
  constraints are expressed as:
  \begin{equation}
    \label{eq:ball-dis}
    \sqrt{ (x_{1} - x_{2})^{2} + (y_{1} - y_{2})^{2} + (z_{1} - z_{2})^{2} } \ge 2r
  \end{equation}
  , where $ r $ is the radius of the two balls, and $ x_{i}, y_{i}, z_{i}, i \in
  \{1, 2\} $ are the 3D coordinates of ball $ i $.
\end{example}

\parh{Two Sources of Invalid Seeds.}~Performing simulations over a physically
infeasible initial state, e.g., a ball penetrating another ball, would also likely lead
to an invalid final state.
We clarify two sources of invalid seed initial states: \ding{192} seed generation
without respecting the constraints; \ding{193} internal bugs in \eng. Oracle
violations due to \ding{192} should be deemed as false positives (FPs).
In contrast, wrong simulation results due to \ding{193} are true positives (TPs), since the invalid seeds are due to bugs in the
tested \eng. We should avoid invalidity issues from \ding{192}.

\parh{Invalidity of Random Seed.}~ Our preliminary experiments show that
randomly generating seed states could easily lead to invalid states from source
\ding{192}. For instance, when arbitrarily generating the initial rotational
angle of a robot arm, we find that a large portion (over 70\%) of the generated
states lead to apparently invalid states, such as the palm penetrating a hard
surface. Even worse, typical \eng\ often lack a comprehensive ability to detect
and reject simulation requests for invalid initial states. For example, in the Brax
Physics Engine~\cite{freeman2021brax} (a Google product), even though the palm
of a UR5E robot arm is obviously under the surface of the ground, the simulation
engine does not detect and warn the user about the issue of such invalid state,
but continues simulation as usual. However, the invalid seeds are not due to
internal bugs of \eng, but arise from wrongly designed seed generation scheme.

\parh{Naive Method: Generate-then-Check (GTC).}~
A potential approach to ensuring state
validity is randomly generating a seed state and rejecting invalid ones.
However, we emphasize that such a generate-then-check (GTC) approach could not
easily work. The constraints are varied across different physical environments.
Hence, it is challenging to present a unified solution for state validity
checking.
Moreover, the format of constraint equations inside a single physical scenario
may be complex, sometimes not even explicitly representable~\cite{10.1145/2159616.2159661,Kim2019}. 

\parh{Our Solution: Simulate-then-Collect (STC).}~In contrast to GTC, which
checks the validity of a generated seed state and rejects invalid, we propose a
simulate-then-collect (STC) approach. The high-level idea of STC is to start
from an initially valid state $ s_{meta} $ (also called \textit{meta seed},
typically can be chosen to be the default initial state shipped by the \seng\ under
test) whose validity can be guaranteed through manual efforts, and run
\textit{forward simulation} to collect a trace $ tr = s_{meta}, s_{1},
s_{2},\ldots, s_{n} $. The states on the simulation trace $ tr $ would then be
collected and added to the seed pool $ \mathcal{C} $ as the fuzzing seeds. To
enhance the diversity of generated seeds, the external forces $ \tau_{i} $ during the transition from state $ s_{i - 1} $ to $ s_{i} $ are varied across different $ i, i \in [1, n] $.

Formally, to see why the STC approach would generate valid states, we present
the following theorem:

\begin{theorem}[Theorem of State Validity]
  \label{theo:validity}
  Assuming the time stepping function $ TS $ of a \seng\ is consistent with
  physics laws, any reachable state $ s_{i} $ starting from an initial
  \textit{valid} state $ s_{meta} $ should also be valid. 

\end{theorem}

\cref{theo:validity} can be easily proved by math induction (details are on \cite{artifact}). We admit that when the time stepping function $ TS $ is
buggy, the collected seed states could be invalid in the first place. However, this should not be a
concern, as to our observation, invalid seeds lead to incorrect simulation
results (true positive bugs), which will be detected by \tool\ later.

\section{Implementation}
\label{sec:implementation}

\tool\ is implemented in about 4K LOC~\cite{artifact}. The key hurdle
of our implementation is accommodating the large differences between evaluated
\eng\ (see \cref{subsec:exp-setting}). Each \seng\ uses a different
domain-specific language (DSL) to describe its simulated environment and perform
forward simulation and gradient computation. The algorithms under each \seng\
are also distinct from each other. Still, our testing method is
\textit{agnostic} to the underlying details of the tested \eng. Moreover, we
spent non-trivial efforts to refactor our code so that extending our testing
method to a different testing configuration requires little engineering effort.
See our codebase and documentation (including instruction on extension)
at~\cite{artifact}.

\section{Evaluation}
\label{sec:evaluation}

In this section, we aim to answer the following research questions (RQs):
\textbf{RQ1}: Can \tool\ effectively uncover errors in the modern \eng?
\textbf{RQ2}: To what extent can the seed scheduling algorithm facilitate the
fuzzing campaign? How much overhead does the scheduling process incur?
\textbf{RQ3}: What are the characteristics of errors uncovered by \tool?
\textbf{RQ4}: What root causes lead to the failures of the tested \eng?
We answer the four RQs from four aspects, respectively: 

\begin{enumerate}[leftmargin=*]
  \item We evaluate our testing methods on eight configurations in total,
including four popular \eng\ and two physical scenarios per \seng. We show that our testing campaign can fruitfully uncover error-triggering inputs on each setting.
  \item We use ablation study to show that our scheduling algorithm
  introduced in \cref{subsec:scheduling} can significantly boost error-discovery
  efficiency while incurring negligible overhead.
  \item We categorize all of our discovered error-triggering inputs to facilitate understanding our findings.
  \item We launch root cause analysis towards our findings through manual
  analysis and developer feedback. We further give a representative example for
  each of our tested \eng\ on the bug root cause.
\end{enumerate}

\subsection{Experiment Setting}
\label{subsec:exp-setting}

\parh{Evaluated \eng.}~Developing \eng\ with both forward and backward
simulation capabilities has been a hot topic in recent years, and numerous
\eng~\cite{geilinger2020add,huang2021plasticinelab,heiden2021neuralsim,de2018end,howell2022dojo}
emerged in just a short period; covering so many existing \eng\ can be
impractical. We carefully review existing \eng\ and choose our testing
targets based on their quality, usability, and popularity. We also consider the
diversity of testing targets, covering \eng\ from both academy and industry.
Ultimately, we select four \eng\ as our evaluation targets, namely, Taichi
Graphics~\cite{hu2019taichi}, Nimble~\cite{werling2021fast}, Warp~\cite{warp},
and Brax~\cite{freeman2021brax}. Nimble is a popular PSE developed by 
Stanford, while Warp and Brax are 
NVIDIA and Google's products, respectively. Taichi Graphics 
originates from MIT researchers and later
transforms into a 50M-startup company, with over 22.4K stars~\cite{taichi-github} on
GitHub.

\parh{Physical Simulation Scenarios.}~We choose two categories of scenarios to
evaluate our tested \eng: one that is fundamental in physical simulation and
generally supported by existing \eng, and the other that is specific to each of
our tested \eng.
For the first category, we choose the simulation of the behavior of a group of
balls bouncing between walls, as the simulation of balls under collision is a
base stone for building up complex simulation cases, such as molecular
simulation~\cite{michaud2011mdanalysis,frenkel2001understanding,sadus2002molecular}
and fluid
simulation~\cite{muller2003particle,amada2004particle,kadau2010atomistic}. Also,
as a fundamental simulation scenario, ball collision is widely supported by
existing \eng~\cite{billiards,warp-pre}. As for the second category, we select
an engine-specific use case for each PSE (four in total). The details of all
the evaluated combinations of \eng\ and scenarios can be found in \cref{tab:eva-conf}. Our tested \eng\ are at their latest versions by the time of writing (except Nimble, see our discussion later). In total, our evaluation consists
of eight configurations (four \eng\ and two scenarios per PSE). For each
configuration, we generate 10K testing inputs and feed the generated testing inputs into
\eng\ for fuzzing.

\begin{table}
  \scriptsize
  \caption{Evaluated configurations. ``All'' means the all the \eng\ are
  evaluated on the scenario of ``Balls.''}
  \label{tab:eva-conf}
  \vspace{-5pt}
  \centering \resizebox{0.85\linewidth}{!}{
    \begin{tabular}{>{\centering\arraybackslash}m{0.6cm}|>{\centering\arraybackslash}m{0.8cm}|>{\centering\arraybackslash}m{1.1cm}|m{4cm}}
      \toprule
      \textbf{\seng} & \textbf{\seng\ Version} & \textbf{Physical Scenario} & \textbf{Brief Description} \\
      \midrule
      All & NA & Balls & A group of balls bouncing around walls and colliding with each other. \\
      \midrule
      Brax & 0.1.0 & UR5E & A UR5 robot arm that can fetch or deliver items in the factories. \\
      \midrule
      Nimble & 0.8.32 & Catapult & A robot with three joints batting balls. \\
      \midrule
      Warp & 0.7.2 & Snake & A set of rods chained in sequence that looks like a snake. \\
      \midrule
      Taichi & 1.4.1 & DiffMPM & An elastic object jumping and moving forward on the ground by deformation. \\
      \bottomrule
    \end{tabular}
  }
\end{table}

\parh{Hyper-parameter Settings.}~As mentioned in \cref{subsec:fuzzing}, we
use threshold $ \epsilon_{B} $, $ \epsilon_{F} $ and $ \epsilon_{I} $ to check
our testing oracles. We also need to decide the initial perturbation amount of $
\Delta s $ on $ s_{0}^{seed} $. Deciding proper values for these parameters is
quite challenging and requires manual efforts (see our clarification on this in
\cref{sec:discussion}). To do so, we repeat the process of generating 1K testing
inputs under a tentatively-decided parameter value and adjusting the value based
on manual inspection on possible false positives (FPs) of findings. We release
all parameter settings in the codebase~\cite{artifact} to enable reproducing
results and research transparency.

\subsection{RQ1: Overall Effectiveness}
\label{subsec:results}

This section studies \textbf{RQ1}, i.e., the effectiveness of our methods on bug
discovery. To that end, we show the total execution time and the number of
errors we find on each combination of tested \eng\ and physical scenarios.
\cref{tab:results} reports the statistics of our experiment. The eight testing
settings are abbreviated via the first letter of the tested PSE and the physical
scenario, e.g., ``\textbf{TD}'' denotes ``\textbf{T}aichi \textbf{D}iffMPM''.

\parh{Processing Time.}~We execute all experiments on a Ubuntu 18.04 server with
an NVIDIA RTX 2080 Ti GPU ported with CUDA 12.1, an Intel(R) Xeon(R) E5-2683
CPU, and 256GB RAM. The third row, ``Time (hr)'', in \cref{tab:results} reports the
execution time (in terms of the number of hours) of each configuration. Overall, the physical simulations on
\eng\ are slow and take dozens to hundreds of hours to finish, as the simulation process requires applying the time-stepping
function (i.e., \cref{eq:forward-seq}) for dozens to hundreds of steps. DiffMPM (\textbf{TD}) is
the most time-consuming one since its
simulation requires discretizing the space into thousands of small grids. The
computation of Balls is lightweight and only takes several hours to finish. An
exception is Brax/Balls (\textbf{BB}), possibly due to its language constraints of disallowing
in-place modifications of array elements~\cite{thinking-in-jax} and thus wasting
time on allocating new arrays for each modification operation.

\parh{Discovered Errors.}~We generate 10K testing inputs for each tested
combination of \eng\ and physical scenarios. The second row, ``\#Errors'', in \cref{tab:results} lists the number of erroneous inputs discovered in each tested setting. Although all of our
tested \eng\ are developed by industry giants or highly experienced researchers
from academia, we can still reveal a considerable number of errors. In
total, we find 5,932 testing inputs that can trigger misbehaviors in the
simulation outputs. 
Specifically, the “Balls” scenarios on Brax and Taichi physics engines have the
highest number of erroneous testing inputs (over a thousand). Although the
physical laws governing the behavior of balls are conceptually simple, the
computer simulation of balls is challenging to be implemented flawlessly (see a
discussion of the author of Taichi on one of the well-known
challenges~\cite{hu2019difftaichi}), and the intricate interactions between
balls and walls may manifest the hidden bugs and thus be captured by our testing
method. Overall, our method uncovers at least 400 erroneous cases in all
testing settings, demonstrating its effectiveness.

\parh{Discovered Crashes.}~Although our testing method primarily focuses on logic
bugs, i.e., bugs that produce incorrect results, we also find several
crashes that cause severe consequences like segmentation fault during our testing campaign.
When a crash happens, we start a new process and continue the generation and
execution of testing inputs using a different random number. 
During the simulation of DiffMPM on Taichi, we find 23 crashes that lead to invalid memory access in GPU. We also
encounter a crash in Nimble that constantly produces segmentation fault and stops the testing process altogether.
The crash is due to a regression bug in the Nimble
physics engine. We thus resort to a relatively older version (0.8.32) of Nimble.

\begin{tcolorbox}[size=small]
\textbf{Answer to RQ1:} \tool\ can effectively find erroneous inputs that
trigger incorrect results or crashes in simulating different physical scenarios
and with various \eng.
\end{tcolorbox}

\begin{table}
  \centering
  \caption{Results overview (10K inputs per testing setting). We leave discussion on ``SS (min)'' in RQ2.}
  \label{tab:results}
  \vspace{-5pt}
  \resizebox{\linewidth}{!}{
    \begin{tabular}{c|c|c|c|c|c|c|c|c|c}
      \toprule
      & \textbf{BB} & \textbf{BU} & \textbf{NB} & \textbf{NC} & \textbf{TB} & \textbf{TD} & \textbf{WB} & \textbf{WS} & \textbf{Total} \\
      \midrule
      \textbf{\#Errors} & 1183 & 475 & 709 & 833 & 1128 & 455 & 427 & 732 & 5932 \\
      \textbf{Time (hr)} & 28.2 & 23.4 & 3.8 & 5.5 & 1.5 & 104.1 & 1.5 & 24.0 & 192 \\
      \textbf{SS (min)} & 65 & 43 & 6 & 2 & 2 & 10 & 4 & 5 & 168 \\
      \bottomrule
    \end{tabular}
  }
\end{table}

\subsection{RQ2: Effectiveness and Overhead of Seed Scheduling}
\label{subsec:ablation}

This section aims to answer RQ2, i.e., the effectiveness of the guided seed
scheduling algorithm in \cref{alg:feedback}, and its incurred time overhead.
To answer this question, we perform an ablation study by excluding the seed
scheduling algorithm in our testing pipeline.

\begin{figure}[htbp]
    \centering
    \includegraphics[width=\linewidth]{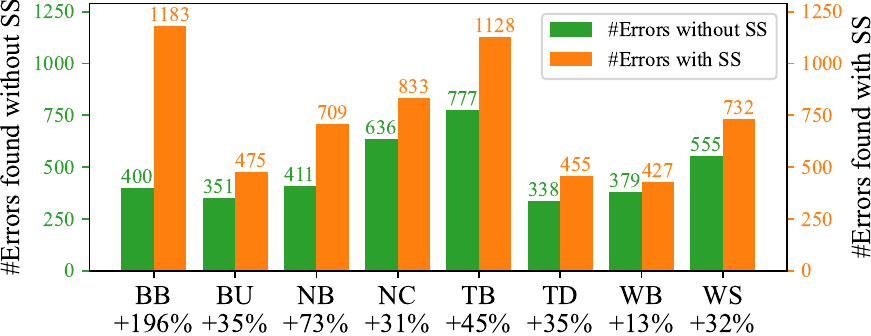}
    \caption{\#Errors discovered without and with seed scheduling (SS). The relative increase (\%) on \#errors of SS is on x-axis.}
    \label{fig:ablation}
\end{figure}

\parh{Ablation Study.}~We disable the seed scheduling algorithm and launch
fuzzing with \textit{exactly the same setting} as \cref{subsec:results}. We also
record the time spent executing the seed scheduling algorithm itself.
\cref{fig:ablation} shows the comparative results between enabling seed
scheduling (abbreviated as ``SS'' in the figure) and disabling it. The green and
yellow bars show the number of errors discovered without and with SS,
respectively. The meaning of first row of labels on the x-axis is the same as \cref{tab:results}, showing all the eight testing combinations; the second row shows the relative
percentage of increase in the number of errors uncovered with SS compared to
without SS.

Overall, the seed scheduling method is seen to significantly boost the error-finding ability. The scheduler typically results in a 30 to 40 percent rise in the
number of uncovered errors, with the highest increase being 195\% and all
exceeding 13\%. The increase is most visible when testing the scenario of ``Balls'' on Brax
(\textbf{BB}) and Nimble (\textbf{NB}) engines. 
Through a later manual analysis on the root cause analysis of the
error-triggering inputs, we find that the conditions to trigger bugs in those
two testing settings are quite intricate. For instance, unveiling some errors
requires multiple objects colliding with each other simultaneously. Such strict
requirements on the error-triggering condition render the random algorithm
struggling in error discovery. In contrast, the seed scheduling algorithm can
prioritize seeds that are more likely to satisfy such conditions.

\parh{Time Overhead of Seed Scheduling.}~Despite the effectiveness of our seed
scheduling algorithm, we find that it only adds negligible amount of time to
the total execution time of fuzzing. The last row, ``\textbf{SS (min)}'', of
\cref{tab:results}, shows the extra time overhead (in terms of minutes) of
seed scheduling (abbreviated as ``SS''). In most cases, the time spent on the
seed scheduling algorithm is within several minutes. In contrast, the total
execution time of the whole fuzzing campaign, as also presented in the row
``Time (hr)'' in the same table, is typically hours to dozens of hours, far
exceeding the time spent on the seed scheduling. The efficiency of the seed
scheduling algorithm partly comes from its succinctness; also, the algorithm
itself can be implemented in terms of simple vector computation primitives, such
as dot products, which existing vector libraries already optimize. 
An exception is the Brax engine, spending around or more than an hour on the
scheduling algorithm. The reason might be similar to its long execution time, as
explained in \cref{subsec:results}. However, compared to the total amount of
time on the simulation (more than 20 hours), the percentage of the overhead
(3\%) is still low.

\begin{tcolorbox}[size=small]
\textbf{Answer to RQ2:} The seed scheduling algorithm can significantly boost
the error discovery while incurring negligible time overhead.
\end{tcolorbox}

\subsection{RQ3: Characteristics of the Errors}
\label{subsec:root-cause}

This section focuses on RQ3.
To answer this question, we utilize a set of heuristics to classify our findings
and summarize some patterns and the distribution of our discovered errors.

\parh{Error Categories.}~As mentioned in \cref{sec:overview}, our testing method
asserts both the forward and backward oracles. Furthermore, forward and
backward errors may exhibit various patterns. We define the following heuristics
to classify the errors on a finer scale:

\parhs{Position \& Velocity Error of System:}~The forward simulation process
essentially deals with computing \textit{states} (see \cref{eq:forward-seq}),
which uniquely defines the system's current situation under simulation. As
introduced in \cref{sec:background}, a state comprises the \textit{velocity}
part and \textit{position} part. We thus characterize the forward errors by
whether the position or the velocity is incorrectly computed during the forward
simulation.

\parhs{Direction \& Extent Error of Gradient:}~The backward simulation computes the gradients w.r.t system initial states. A gradient is a vector with \textit{direction} and \textit{extent}. Accordingly, we classify backward errors
into direction errors and extent errors.

\parh{Criteria for Error Classification.}~When classifying forward errors,
we compare the mutated initial state $ s_{0}^{mut} $ with its corresponding seed
initial state $ s_{0}^{seed} $, and deem it to be a position (velocity) error if
their position (velocity) components of the state significantly deviate. As for
determining whether the gradient direction is wrong for a backward error, we use
the following heuristic:
\begin{equation}
  \label{eq:dir-err}
  \arccos \frac{-g(\Delta s)^{T}\Delta s}{\lVert g(\Delta s)\rVert \lVert \Delta s \rVert} < \theta_{D}
\end{equation}
, where $ \Delta s = s_{0}^{mut} - s_{0}^{seed} $, $ g(\Delta s) $ is the
gradients computed by the PSE w.r.t. $ \Delta s $, and $ \theta_{D} $ is a
threshold, which is empirically set to $ \pi / 2 $. \cref{eq:dir-err} asserts that the gradient vector $ g(\Delta s) $
should point to the direction close to $ -\Delta s $. A violation of
\cref{eq:dir-err} indicates erroneous gradient direction computed from \seng.

In terms of classifying the gradient extent errors, we use the following heuristic:
\begin{equation}
  \label{eq:ext-err}
  \forall || h(\Delta s_{1}) || > || h(\Delta s_{2}) ||, 
  || g(\Delta s_{1}) || \geq || g(\Delta s_{2}) ||
\end{equation}

Intuitively, \cref{eq:ext-err} asserts that a $ \Delta s $ leading to smaller loss value $ h(\Delta s) $ should not
receive a larger gradient extent. Such heuristic is based on a general
observation that the optimization process should gradually converge as the loss
value decreases, hence the gradient extent should decrease accordingly.

\begin{table}
  \centering
  \caption{Classification of error-triggering inputs. ``Unapparent
  errors'' are possibly false positives of \tool, since unlike the rest of cases, they do not have
  apperantly erroneous patterns. See further discussion in \cref{sec:discussion}.}
  \label{tab:classification}
  \resizebox{\linewidth}{!}{
  \begin{tabular}{ c|c|c|c|c|c|c|c|c|c }
      \toprule
      & & \textbf{BB} & \textbf{BU} & \textbf{NB} & \textbf{NC} & \textbf{TB} & \textbf{TD} & \textbf{WB} & \textbf{WS} \\
      \midrule
      \multirow{2}{*}{\textbf{Forward}} & \textbf{Position} & 145 & 227 & 88 & 258 & 92 & 44 & 89 & 142 \\
       & \textbf{Velocity} & 139 & 208 & 79 & 384 & 130 & 41 & 73 & 130 \\
      \midrule
      \multirow{2}{*}{\textbf{Backward}} & \textbf{Direction} & 669 & 71 & 452 & 395 & 874 & 300 & 360 & 420 \\
       & \textbf{Extent} & 679 & 108 & 156 & 271 & 515 & 370 & 205 & 375 \\
       \midrule
       \multicolumn{2}{c|}{\textbf{Unapparent Errors}} & 156 & 16 & 67 & 19 & 41 & 23 & 5 & 14 \\
      \bottomrule
  \end{tabular}
  }
\end{table}

\parh{Classification Results.}~\cref{tab:classification} shows the number of
errors in each category for all the tested \eng\ and physical scenarios. 
Note that the sum of all four categories does not equal the total number of
errors listed in \cref{tab:results}, because an error-triggering input may
exhibit multiple patterns, such as being incorrect in both the gradient
direction and extent.

Overall, we discover a considerable number of errors in all four categories.
Based on the statistics, the errors uncovered by \tool\ extensively affect
different phases of \eng. For example, during the forward simulation, the
simulated objects may deviate from their correct destination when the position
changes, and the objects may unexpectedly move at varying speeds when the
velocity changes. Similarly, in the backward phase, the gradient direction
errors mislead the optimization direction, and the optimization process fails to
converge in the existence of gradient extent errors. Our exposed diverse
errors, in turn, can lead to various harmful consequences, ranging from objects
appearing in confusing positions or moving at noticeably unreal speeds, to the
failure to achieve desired goals in agent learning and control tasks. For
instance, one of the discovered errors has already caused major confusions to
the community when users are performing downstream tasks~\cite{brax-bug}. Due
to limited space, we present all our findings on our website~\cite{artifact} for users to re-produce and extend.

We also find that some of our discovered erroneous inputs do not fall into any
of our listed classification criteria (the row ``Unapparent Errors'') in
\cref{tab:classification}. Since they do not have obviously erroneous
patterns, we conservatively deem them as possible false positives of \tool. This
is likely due to hyper-parameter settings (see our discussion in \cref{sec:discussion}).

\begin{tcolorbox}[size=small]
\textbf{Answer to RQ3:} Our uncovered erroneous test cases feature a wide range
of characteristics, covering position and velocity errors in the forward
simulation, as well as gradient direction and extent errors in the backward
process.
\end{tcolorbox}

\subsection{RQ4: Root Cause Analysis}
\label{subsec:root-cause}

This section looks into RQ4: what are the root causes of discovered failures? To
further the understanding of discovered error-triggering inputs, we pack all of
our findings to the developers of tested \eng. With developers' help and
valuable feedback, we have rooted seven bugs by the time of writing, two of
which are promptly fixed in the recent release. We clarify that it takes some time for
developers to iterate all uncovered errors since root cause analysis requires
special expertise, and the stealthiness nature of logic errors and the large
number of buggy cases (hundreds or even a thousand) increase the analysis difficulty.
For illustrative purposes, below we provide a representative example for each tested
\eng.

\noindent\textit{\textbf{A Bug in Taichi:}~Incorrect Compilation of Gradient
Computation Code.}~The development of \eng\ is a highly complex task, requiring
not only programming the simulator, but sometimes also developing a
compiler that can generate highly efficient executables for the code of the
simulator. The Taichi engine has its DSL, the Taichi-lang, and a dedicated
compiler that can generate executables for code written in its DSL. We find that
\texttt{ti.static}, a primitive in Taichi-lang that can be used as an indicator
for loop unrolling, causes incorrect outputs in the computed gradients. In
particular, the compiler in Taichi would automatically generate code that
computes the analytical gradients, so that developers of simulators only need to
focus on programming the forward simulation and can leave the gradient
computation to the compiler. However, the compiler of Taichi, although correctly
compiles the forward simulation code and gives correct results in the forward
simulation, mis-compiles the semantics of \texttt{ti.static} and generates
incorrect executables for gradient computation.

\noindent\textit{\textbf{A Bug in Nimble:}~Improper Handling on Impulse
Computation.}~When solving the dynamics equation, as the example provided in
\cref{eq:newton-law} and \cref{eq:acceleration}, some \eng, such as Nimble,
would first compute the \textit{impulses} applied on the objects, i.e., the
product of a time interval $ \Delta t $ with force on the corresponding object.
The impulse would later determine the change of objects' velocities in the given
time interval $ \Delta t $. The impulse computation of Nimble is correct in most
situations, yet we find that when multiple objects collide with each other
simultaneously, the output impulses are wrong. Nimble employs a Linear
Complementarity Problem (LCP) solver to compute the impulses; however, the LCP
solver neglects the special case of simultaneous collision. The impulses are
applied twice on each pair of the colliding objects; thus, the velocity changes
twice as much as it should. Consequently, the total momentum of the whole system
is doubled, a phenomenon that obviously violates the law of momentum.

\noindent\textit{\textbf{A Bug in Brax:}~Erroneous Contact Modeling.}~Contacts
between objects are prevalent in both the physical world and simulations. For
instance, a walking robot must have its feet in contact with the ground to
perform the ``walking'' movement~\cite{hu2019difftaichi}, and a ball bouncing
off a wall would create instant contact between the ball and the wall. Brax
models object contacts using the Position Based Dynamics
(PBD)~\cite{muller2007position}. Recall in \cref{sec:background}, \eng\ would
discretize the time into small-time intervals $ \Delta t $. Brax's PBD implementation incorrectly models the objects' contact \textit{only at the end
of} the interval $ \Delta t $, even though in reality, two objects may touch
each other at any time \textit{in between} a $ \Delta t $ time span. Such an
error is also independently discovered by a user who wants to use Brax to
perform a downstream task on agent training and learning. The user complains
about poor training results due to the wrong simulation. Developers deem the contact modeling issue as ``definitely one of the big
problems plaguing most simulators''~\cite{brax-bug}.

\noindent\textit{\textbf{A Bug in Warp:}~GPU Kernel Caching and Replaying
Issue.}~Since the physical simulation process inherently requires intensive
computational resources, some \eng\ may use GPUs to accelerate the simulation
process. The Warp Engine launches a set of GPU kernels to execute the forward
and backward simulation. Since a GPU kernel may be launched multiple times
during the simulation, Warp would cache previously launched kernels and replay
them later when the same kernels are called again. Through manual investigation,
we find that the kernel caching mechanism may sometimes cause the break of the
backpropagation flow in the gradient computation phase, resulting in some
updated parameters receiving zero gradients.

\begin{tcolorbox}[size=small]
\textbf{Answer to RQ4:} We find a wide variety of error-inducing root causes in
the whole software stack of \eng, spanning from simulation algorithm design, compiler, and hardware support. Such
observation indicates the generalizability of our testing method and the
importance of our findings.
\end{tcolorbox}

\section{Discussion}
\label{sec:discussion}

\parh{Limitations and Threats to Validity.}~Our study launches
\textit{dynamic testing} toward \eng. Although our method successfully uncovers
a considerable number of errors, we cannot \textit{guarantee} the correctness of
\eng. Program verification may ensure functional correctness, yet it is
generally considered extremely challenging to implement for complex software
like \eng. Overall, our testing-based approach is in line with the general
stance of relevant works~\cite{sun2016finding,le2015finding}.

Another limitation is the generalizability of our testing oracles. We enforce a
set of assumptions in our testing oracles (see \cref{sec:overview}). For other
physical scenarios, such as the Lorenz System~\cite{MOGHTADAEI2012733}, in which
our assumptions do not hold, our testing method cannot be directly applied to check the
correctness of simulations. Still, our testing method is seen
to be promising as it successfully uncovers a considerable number of errors in
all the tested, mainstream \eng.

\parh{Dicussion of Hyper-parameter Settings.}~Overall, we have two sets of
hyper-parameters: the first set is perturbation amount $ \Delta s $ on the seed
state, and the second is $ \epsilon_{F} $, $ \epsilon_{I} $, and $ \epsilon_{B}
$ for asserting the testing oracles.
Poorly-decided hyper-parameters may lead to false positives (FPs).
However, deciding a reasonable value setting requires expertise about the tested \seng\ and physical scenario, and cannot be easily automated, since different combinations of \seng\ and scenario have varied features in their simulation process. For instance, the softbody simulations simulate thousands of small particles and grids~\cite{hu2019taichi}, while robotic simulations are concerned with dozens of complex-shaped components, e.g., palms and shoulders~\cite{werling2021fast}; also, different \eng\ may leverage different algorithms for the simulation. Without in-depth understanding of the features of the \eng\ and scenarios, the selected hyper-parameter values may not be reasonable under the given setting. As such, it is challenging to design an automated, unified method for the parameter decision process.
We thus adopt an empirical approach, following the manual investigation process as mentioned in \cref{subsec:exp-setting}.
To make our results fully transparent and highly reproducible, we have
released all of our selected hyper-parameters in our codebase.

\parh{Discussion about Alternative Testing Oracles.}~Since there exist many
\eng\ in public repositories, one may question the feasibility of launching
differential testing over a set of \eng\ using an identical initial state. Our
preliminary study illustrates that such differential testing setting is
problematic, whose reasons are two-fold.

\parhs{Different Modeling of Physics Laws.}~\eng\ may use different models for
the simulated physical process. Consider the simulation of continuum dynamics,
some may use the St. Venant-Kirchhoff model~\cite{10.1145/1073204.1073300}, while others may use
the corotated linear elasticity for modeling the simulated physical
process~\cite{sifakis2012fem}. The two models lead to different 1st
Piola-Kirchhoff stress tensors~\cite{CHEN2019177}, thus the final results may
be different. As such, the results from the two engines are not comparable.

\parhs{Different Equation-Solving.}~Forward simulation in \eng\ may give
different results due to different equation-solving schemes. The forward
simulation process would first discretize the dynamics equations, yet the
discretization schemes vary from engine to engine. The implicit Euler method will dampen the
Hamiltonion~\cite{10.1145/1185657.1185669} of the simulated system, while the symplectic Euler
method preserves it~\cite{10.1145/1185657.1185669}. Thus, the forward simulation
results from different engines may not be consistent; the computed gradients, as
a consequence, can also be inconsistent.

\vspace{-2pt}
\section{Related Work}
\label{sec:related}

Our research focuses on discovering bugs in physical simulators, which are software systems that model real-world physical phenomena. While there is a body of work~\cite{10.1145/3597926.3605233,10.1145/3180155.3180231,10.1145/3463274.3463806,10.1145/3377812.3382163,9283988,10.1145/3540250.3549159} that tests the Simulink compiler, a compiler for multidomain dynamical systems, our work is distinct and orthogonal to those efforts. Simulink models are user-defined and may not adhere to real-world physical laws, whereas our work concentrates on testing physical simulators that mimic real-world systems and relies on a set of established physical principles.

Our approach involves generating a series of random testing inputs to stress-test our testing targets, which are the physical simulators. This method, known as fuzzing, has been applied to test other targets as well, such as Markov decision processes~\cite{10.1145/3533767.3534388} and decompilers~\cite{10.1145/3395363.3397370}. Additionally, fuzzing has been used to identify missed optimizations in WebAssembly compilers by randomly generating C programs~\cite{10.1145/3597926.3598068}.
While fuzzing creates new input data to test software systems, another approach called metamorphic testing transforms existing input data to uncover bugs.
Metamorphic testing has been applied to test complex software systems, including Deep Learning compilers~\cite{10.1145/3489048.3522655}, graphics shader compilers~\cite{10172737,10.1145/3133917}, AI models~\cite{10.1145/3551349.3561157}, and database systems~\cite{10.14778/3494124.3494139}. In contrast, our approach focuses on using fuzzing to test physical simulators, which presents unique challenges and opportunities for bug detection.

\section{Conclusion}

We present a novel fuzzing framework, \tool, targeting modern \eng. \tool\ is
designed on the basis of principled physical laws to uncover logic errors in
\eng, and it features a set of design choices and optimizations to improve the
testing. Our study on eight combinations of \eng\ and physical scenarios
detects a considerable number of findings, covering the full stack of \seng\
software. This work can serve as a roadmap for users and developers to use
and improve \eng.

\section*{Acknowledgement}

We thank anonymous reviewers for their valuable feedback. This project is 
supported in part by a RGC ECS grant under the contract 26206520.

\bibliographystyle{IEEEtran}
\bibliography{bib/machine-learning,bib/timing,bib/sidechannel,bib/analysis,bib/ref,bib/testing-cv,bib/cv,bib/sw}

\end{document}